\documentclass[         %
aps,                    
prd,                    
showpacs,               
nofootinbib,            
twocolumn,             %
showkeys,               %
preprintnumbers,        %
floatfix]               
{revtex4}               
\usepackage{graphicx,longtable}
\usepackage{dcolumn}

\def\Red  {}
\def\BrightRed  {}
\def\Black{}
\def\Green{} 
\def\Blue {}

\begin{document}
\title{\BrightRed Do the KamLAND and Solar 
Neutrino Data Rule out Solar Density Fluctuations?}
\author{\Blue        A.~B. Balantekin\Black}
\email{         baha@nucth.physics.wisc.edu}
\author{\Blue        H. Y\"{u}ksel}
\email{         yuksel@nucth.physics.wisc.edu}
\affiliation{\Green  Department of Physics, University of Wisconsin\\
               Madison, Wisconsin 53706 USA \Black}
\date{\today}
\begin{abstract}
\Red We elucidate effects of solar density fluctuations on neutrino
propagation through the Sun.  Using data from the recent solar
neutrino and KamLAND experiments we provide stringent limits on solar
density fluctuations. It is shown that the neutrino data constrains
solar density fluctuations to be less than $\beta = 0.05$ at the 70 \%
confidence level, where $\beta$ is the fractional fluctuation   around
the value given by the Standard Solar Model. We find that the best fit
to the combined solar neutrino and KamLAND data is given by $\beta =
0$.  \Black
\end{abstract}
\medskip
\pacs{14.60.Pq, 26.65.+t, 96.60.Jw, 96.60.Ly} 
\keywords{Solar Density Fluctuations, Solar
Neutrinos, Reactor Neutrinos} 
\preprint{} 
\maketitle

\Blue
\section{Introduction}
\Black

The evolution of the Sun is modeled using a standard set of input
parameters describing the physics of the interior, including
thermodynamic properties (equation of state), energy transfer through
solar matter (opacity) and the rates of the nuclear reactions that
power the Sun (astrophysical S-factors). The resulting detailed models
of the Sun predict temperature, density, composition profiles, and
neutrino fluxes coming from the  nuclear fusion reactions. Recently
the Standard Solar Model (SSM) has  been enjoying a tremendous
success. Its predictions (see \cite{Bahcall:2000nu,Brun:1998qk}) have
withstood the observational and experimental tests very well. For
example, SSM predicts frequencies of the pressure (p-mode) vibrations
which can be observed on the solar surface. Observation and analysis
of these oscillations, called helioseismology, provide detailed
information about the Solar interior (for a thorough introduction to
helioseismology and review of the observational data see Ref.
\cite{Christensen-Dalsgaard:2002ur}). The helioseismological
information can be inverted to obtain  the sound-speed profile
throughout the Sun. The speed of sound in the Sun is determined by
combining the density and temperature profiles which are predicted
using the Standard Solar Model. The sound speeds of solar models that
include element diffusion agree with helioseismological measurements
to better than 0.2 \% \cite{Bahcall:1996qw,Bahcall:1998wm}.  Another
test of SSM was achieved by the neutral-current measurements  at the
Sudbury Neutrino Observatory (SNO) \cite{Ahmad:2002jz}. These
measurements  yield a total (all flavors) $^8$B solar neutrino flux
which is in very good agreement with the Standard Solar Model
predictions. Identifying the precise nuclear reactions that power the
Sun would be another valuable constraint on the SSM
\cite{Adelberger:1998qm}. It is generally believed that the nuclear
fusion reactions that power the Sun take place in the so-called pp
cycle. The recent data also provides a stringent limit (less than 7
\%) to the amount of energy that the Sun produces via the CNO fusion
cycle \cite{Bahcall:2002jt}.

Given these recent successes of the Standard Solar Model and the
quality of the experimental data currently being taken, perhaps the
time has come to test some of the other aspects of the model. One such
test, namely looking for fingerprints of the solar density
fluctuations in solar neutrino spectra was already considered
extensively starting in the early 1990's
\cite{Loreti:1994ry,Balantekin:1996pp,Nunokawa:1996qu,Burgess:1996mz,%
Nunokawa:1997dp,Bamert:1997jj,Torrente-Lujan:1998pf,Bykov:2000ze}, but
at that time the solar  neutrino data were not accurate enough to make
a definitive statement. Such fluctuations may provide an additional
probe of the physics of the deep solar interior
\cite{Burgess:2002we}. For example, evolution theories of stars, in
addition to p-modes, also predict the existence of buoyancy-driven
gravity modes (g-modes). Whether g-modes are actually excited in the
Sun is an open question. These modes are exponentially damped in the
convective zone of the Sun; unlike the p-modes it is not possible to
observe the resulting very small amplitudes of the g-modes on the
surface of the Sun using current techniques even if they are indeed
excited in the Sun.  It is argued that g-modes in the Sun can be
excited by turbulent stresses in the convective zone
\cite{Kumar:1995gq} or by magnetic fields in the radiative zone
\cite{Burgess:2002we}. Although the latter hypothesis requires large
magnetic fields in the Sun, a magnetic field as large as $10^7$ G in
the radiative zone seems to be permitted by the helioseismic data
\cite{Couvidat:2002bs}. It should be emphasized that temperature
fluctuations associated with the large-amplitude g-mode oscillations
would have significantly reduced the neutrino flux {\em produced} in
the core of the Sun. Observational upper limits on the surface
velocity amplitudes (as well the direct measurement of the total solar
neutrino flux at SNO) rule out such large-amplitude g-mode
oscillations \cite{Bahcall:zk}. The scenario we discuss here is the
possibility of smaller-amplitude g-mode oscillations or some other
mechanism causing fluctuations in the density profile of the Sun
affecting neutrinos only through their {\em interaction} with the
matter. (For an alternative mechanism for producing density
fluctuations in the Sun through magneto-hydrodynamic waves see
Ref. \cite{Dzhalilov:1998np}).  At the very least it is important to
investigate what limits the solar neutrino data (supplemented by the
constraints of the KamLAND reactor neutrino measurements) would place
on the size of such fluctuations.

Our goal in this paper is to revisit the subject of density
fluctuations and investigate limits  placed by the recent solar and
reactor neutrino experiments on the amount of fluctuation.  Our 
preliminary attempts to use earlier SNO data were presented in
Ref. \cite{Balantekin:2001dx}.  In the next section we review the
formalism and summarize our method for doing the global analysis. In
Section 3 we present our results and discuss their implications.

\Blue
\section{Solar Density Fluctuations}
\Black

We will assume that the electron density $N_e$ fluctuates around the
value, $\langle N_e \rangle$, predicted by the SSM
\begin{equation}
N_e (r) = (1 + \beta F (r)) \langle N_e (r) \rangle , 
\label{flucdef}
\end{equation}
and that the fluctuation $F (r)$ obeys
\begin{eqnarray}
\langle F (r) \rangle & = & 0 \nonumber \\
\langle F (r_1) F (r_2) \rangle & = & f_{12} \nonumber \\ 
\langle F (r_1) F (r_2) F (r_3) \rangle & = & 0 \nonumber \\
\langle F (r_1) F (r_2) F (r_3) F (r_4)\rangle & = &  
        (f_{12} f_{34} + f_{13} f_{24} + f_{14} f_{23}) \nonumber\\
& \vdots   &
\label{eq:fluct-cond}
\end{eqnarray}
where $f_{ij} = f(|r_j - r_i|)$ gives the correlation between
fluctuations in different places. In Eq. (\ref{flucdef}) we can
interpret the quantity $\beta$ as the fraction of the fluctuation
around the density given by the SSM. Throughout the current paper, we
consider the case of delta-correlated (white) noise:
\begin{equation}
f(r) = 2 \tau \delta (r),
\label{wn}
\end{equation}
with the correlation length $\tau$ as a parameter. We discuss the
limitations of this assumption later in this section. The Hamiltonian
describing neutrino evolution in matter can be written as as sum of
two terms:
\begin{equation}
H = H_0 + B(r) M,
\label{eq:1}
\end{equation}
where $H_0$ is the standard Mikheev-Smirnov-Wolfenstein 
(MSW) Hamiltonian \cite{Wolfenstein:1977ue} (for a brief review see 
\cite{Balantekin:1998yb}) 
with the average electron
density $\langle N_e \rangle$, $B(r)$ is the fluctuating c-number and
$M$ is a constant operator. For the two-flavor mixing $H_0$ 
governs the time-evolution in the SSM density: 
\begin{eqnarray}
i \frac{\partial}{\partial t}
 \left( \begin{array}{c} \nu_e \\ \nu_x \end{array} \right) &=&
\frac{\delta m^2}{4 E}
\left( \begin{array}{cc}
\zeta(t) - \cos 2\theta & \sin 2\theta \\
\sin 2\theta            & -(\zeta(t) - \cos 2\theta)
\end{array} \right) \nonumber \\
&& \times
\left( \begin{array}{c} \nu_e \\ \nu_x \end{array} \right)
\label{eq:msw}
\end{eqnarray}
where
\begin{equation}
\zeta(r) = \frac{2 \sqrt{2} G_F E}{\delta m^2} \langle N_e(r) \rangle,
\end{equation}
$\theta$ is the vacuum neutrino mixing angle, $\delta m^2$ is the
difference in the squared masses of the two neutrino species, $E$ is
the neutrino energy, and $\langle N_e \rangle$ is the averaged
electron number  density given by the SSM. In Eq. (\ref{eq:msw}),
$\nu_x$ is an arbitrary combination of $\nu_{\mu}$ and $\nu_{\tau}$
\cite{Balantekin:1999dx}.  The fluctuating term $B(t)$ is given by
\begin{equation}
B (r) =  \beta F (r) \frac{G_F}{\sqrt{2}} \langle N_e (r) \rangle,
\label{fluc} 
\end{equation}
and $M = \sigma_3$. Using Eq. (\ref{eq:fluct-cond}) one can show
that the fluctuation of $B(r)$ satisfies the conditions
\begin{eqnarray}
\langle B(r_1) \rangle & = & 0 \nonumber \\
\langle B(r_1) B(r_2) \rangle & = & \alpha^2 f_{12} \nonumber \\
\langle B(r_1) B(r_2) B(r_3) \rangle & = & 0 \nonumber \\
\langle B(r_1) B(r_2) B(r_3) B(r_4)\rangle & = & \alpha^4
        (f_{12} f_{34} + f_{13} f_{24} + f_{14} f_{23}) \nonumber \\
& \vdots  &
\label{fluct-cond2}
\end{eqnarray}
where
\begin{equation}
\alpha (r) = - \frac{G_F}{\sqrt{2}} \beta \langle N_e \rangle,
\end{equation}
It was shown in Refs. \cite{Loreti:1994ry,Balantekin:1996pp} that
with the white noise assumption of Eq. (\ref{wn})  the
fluctuation-averaged neutrino density matrix satisfies the equation
\begin{equation}
\frac{\partial}{\partial t} \langle \rho(t) \rangle =
-\alpha^2 \tau [M, [M, \langle \rho(t) \rangle]]
- i[H_0(t), \langle \rho(t) \rangle].
\label{eq:avgev}
\end{equation}
For the two-flavor case, depicted in Eqs. (\ref{eq:msw}) and
(\ref{fluc}), after calculating the commutators Eq. (\ref{eq:avgev})
can be re-written as a $3 \times 3$ matrix equation
\cite{Loreti:1994ry,Balantekin:1996pp}
\begin{equation}
\frac{\partial}{\partial t}
\left( \begin{array}{c} z \\ x \\ y \end{array} \right)
= -2 \left( \begin{array}{ccc}
 0  &     0  &      D \\
 0  &     k  &  -A(t) \\
-D  &  A(t)  &      k
\end{array} \right)
\left( \begin{array}{c} z \\ x \\ y \end{array} \right) , 
\label{eq:meanev}
\end{equation}
where the individual elements of the density matrix are 
\begin{eqnarray}
z &=& 2 \langle \nu_e^* \nu_e^{\phantom{*}} \rangle - 1 \nonumber \\
x &=& 2 \> \hbox{Re} \, \langle \nu_\mu^* \nu_e^{\phantom{*}} \rangle
        \nonumber \\
y &=& 2 \> \hbox{Im} \, \langle \nu_\mu^* \nu_e^{\phantom{*}}
        \rangle.
\end{eqnarray}
In these equations $\nu_f$ is the probability amplitude for the
neutrino flavor $f$, and we introduced the definitions
\begin{equation}
A(t) \equiv \frac{\delta m^2}{4E} (\zeta(t) - \cos
2\theta), \quad D \equiv \frac{\delta m^2}{4E} \sin 2\theta ,
\end{equation}
and
\begin{equation}
k = G_F^2 \langle N_e (r) \rangle^2 \beta^2 \tau.   
\label{kformula} 
\end{equation}

We numerically solved Eq. (\ref{eq:meanev}) using the technique
developed in Ref. \cite{Ohlsson:1999um} which we summarize. We first
consider the constant density case. If we represent
the column vector in Eq. (\ref{eq:meanev}) by ${\cal R}(t)$ and the
matrix governing its time-evolution by ${\cal K}$ ($d{\cal R}/dt =
{\cal K}{\cal R}$), for constant density one can write
\begin{equation}
{\cal R}(t)  = \exp{ [{\cal K} t ]} {\cal R}(t=0) = U (t) 
{\cal R}(t=0) .
\label{ch1} 
\end{equation}
The exponential in Eq. (\ref{ch1}) can be calculated using
Cayley-Hamilton theorem which states that for any matrix $M$,  the
eigenvalue in the characteristic equation can be replaced with the
matrix itself. For a $3 \times 3$ matrix we get:
\begin{equation}
e^M = a_0 I + a_1 M + a_2 M^2 .
\label{ch2} 
\end{equation}
The coefficients can be found by by substituting eigenvalues
$\lambda_i$ of $M$ into
\begin{equation}
e^{\lambda_i} = a_0 +\lambda_i a_1 + \lambda_i^2 a_2.
\label{ch3} 
\end{equation}
In a medium with varying density (where the variables $A$ and $k$ are
changing), time evolution can be calculated by dividing neutrino path
into small intervals in which potential can be approximated as a
constant. Then the total evolution operator is  product  of evolution
operators for all intervals:
\begin{equation}
U = U_N U_{N-1}....... U_2 U_1 .
\label{ch4} 
\end{equation}
Care must be taken to ensure that the intervals chosen are smaller
than the oscillation length.  
 
\begin{figure}[t]
\vspace*{+.3cm}
\includegraphics[scale=.33]{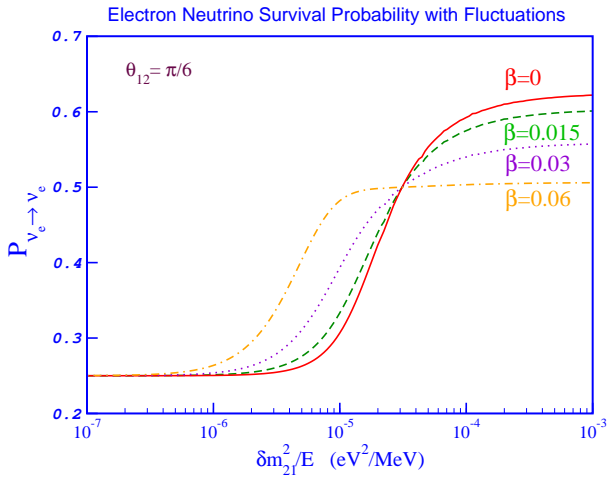}
\vspace*{+0cm} \caption{ \label{fig:1} Mean survival probabilities for
  the SSM density profile and $\sin \theta_{12} = \pi/6$ calculated as
  described in the text. The correlation length is chosen to be 10 km
  and the probabilities are plotted for the percentage fluctuation
  values of $\beta = 0$ (solid line), $\beta = 0.015$ (dashed
  line),$\beta = 0.03$ (dotted line),$\beta = 0.06$ (dot-dashed line).}
\end{figure}

\begin{figure}[t]
\vspace*{+.3cm}
\includegraphics[scale=.33]{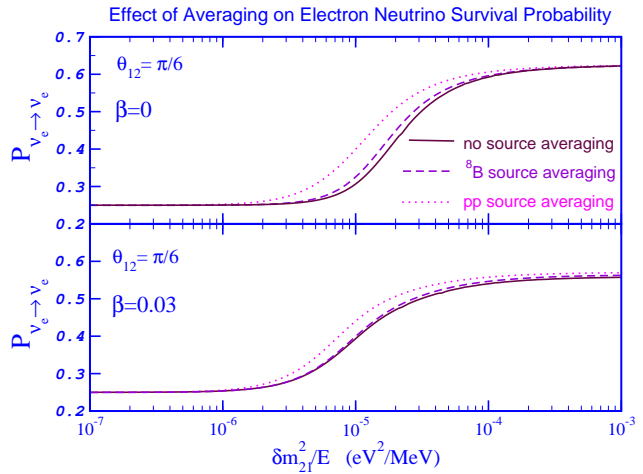}
\vspace*{+0cm} \caption{ \label{fig:2} The effect of source averaging
  on various components of the mean survival probability without
  ($\beta=0$) and with ($\beta=0.3$) fluctuations. The SSM density
  profile of Ref. \cite{Bahcall:2000nu}, the mixing angle of $\sin
  \theta_{12} = \pi/6$, and the correlation length of 10 km are used.
  The solid lines are the survival probabilities without source
  averaging. The dashed lines represent the source-averaged survival
  probabilities of $^8$B neutrinos and the dotted-line those of the pp
  neutrinos.}
\end{figure}

To give a feeling about the solutions of  Eq. (\ref{eq:meanev}) we
show in Figure \ref{fig:1} mean survival probabilities obtained by
solving it numerically as described above for different values of
$\beta$ and a correlation length of $\tau = 10$ km. From this figure
one observes that the effect of the fluctuations is more dominant when
the neutrino parameters and the average density are such that neutrino
evolution in the absence of fluctuations is adiabatic. For the
two-flavor case one can write the electron neutrino survival
probability as \cite{Haxton:dm,Balantekin:1998yb}
\begin{equation}
P (\nu_e \rightarrow \nu_e) = \frac{1}{2} + \frac{1}{2} \cos 2 \theta
\cos 2 \theta_M ( 1 - 2 |\psi_2|^2 ), 
\label{ch5} 
\end{equation}
where $|\psi_2|^2$ is the probability of observing the second matter
eigenstate on the surface of the Sun (calculated with the initial
condition that the neutrinos start in the first matter eigenstate),
also known as the hopping probability, and the matter angle is given
as
\begin{equation}
\cos 2 \theta_M = - A / \sqrt{A^2 + D^2}.
\label{ch6} 
\end{equation}
In Eq. (\ref{ch5}) $\cos 2 \theta_M$ is the matter angle where the
neutrinos are created.  Note that the $\cos 2 \theta_M$ is zero at the
MSW resonance.   If the neutrino propagation is adiabatic when $\beta$
is set to zero, the hopping probability is zero. However when the
fluctuations are turned on they cause some hopping, yielding a small
but non-zero  $|\psi_2|^2$, which grows with $\beta$. Consequently
when $\beta \neq 0$, the absolute value of the second term on the
right-hand side of  Eq. (\ref{ch5}) is always  less than its value
when $\beta = 0$.  If the value of $\delta m^2 / E$ is such that
neutrinos are produced at the MSW resonance density then the cosine of
the initial matter angle is zero, and  Eq. (\ref{ch5}) predicts a
survival probability of $1/2$ no matter what the value of $\beta$
is. For smaller values of $\delta m^2 / E$ the initial value of the
matter angle is negative, yielding a higher survival probability as
compared to the $\beta =0$ case. For larger values of $\delta m^2 / E$
the initial value of the matter angle is positive, yielding a lower
survival probability as compared to the $\beta =0$ case. This behavior
is clearly evident in Figure \ref{fig:1}.

There are two constraints on
the value of the correlation length. In averaging over the
fluctuations we assumed that the correlation function is a delta
function (cf. Eq. (\ref{wn})). In the Sun it is more physical to
imagine that the correlation function is like a step function of size
$\tau$. Assuming that the logarithmic derivative is small, which is
accurate for the Sun, delta-correlations are approximately the same as
step-function correlations if the condition
\begin{equation}
\tau \ll \left( \sin 2\theta \frac{\delta m^2}{2E} \right)^{-1}
\label{eq:step-approx-msw}
\end{equation}
is satisfied \cite{Balantekin:1996pp}. 
Eq. (\ref{eq:step-approx-msw}) can be rewritten as
\begin{equation}
\tau ({\rm km}) \ll 3.95 \times 10^{-4} 
\frac{E ({\rm MeV})}{\sin 2\theta \> \delta m^2 ({\rm eV}^2)} .
\label{tau1}
\end{equation}
A second constraint on the correlation length is provided by the
helioseismology. Density fluctuations over scales of $\sim 1000$ km
seem to be ruled out
\cite{Christensen-Dalsgaard:2002ur,Burgess:2002we,Castellani:1997pk}.
On the other hand current helioseismic observations are rather
insensitive to density variations on scales close to $\sim 100$ km
\cite{Burgess:2002we}. 

\begin{figure}[t]
\vspace*{+.3cm}
\includegraphics[scale=.33]{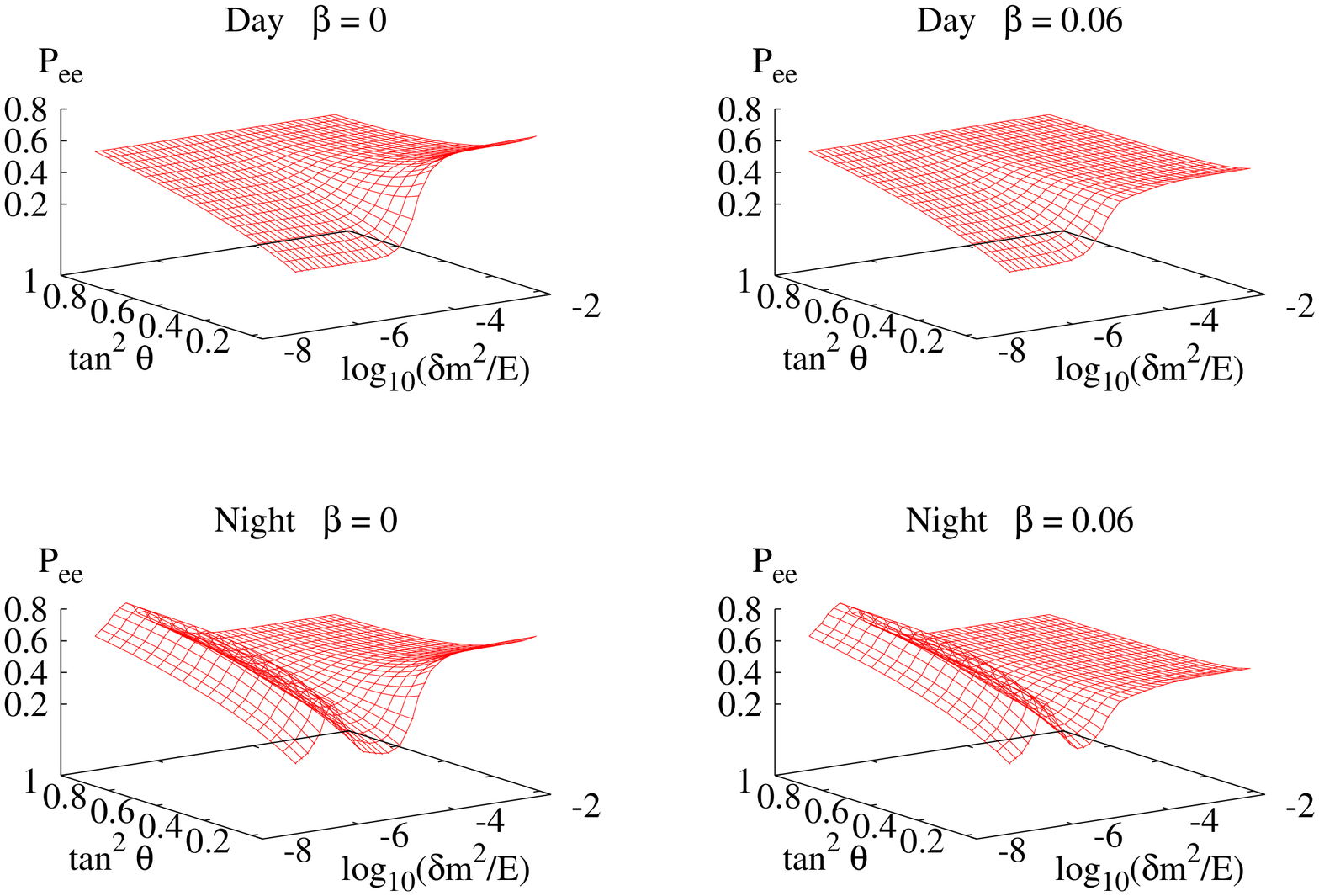}
\vspace*{+0cm} \caption{ \label{fig:3} Source-averaged survival
  probabilities for $^8$B neutrinos with and without solar-density
  fluctuations detected at the location of SNO during the day and the
  night. The SSM density profile of Ref. \cite{Bahcall:2000nu} and the
  correlation length of 10 km are used.}
\end{figure}

\begin{figure}[t]
\vspace*{+.3cm}
\includegraphics[scale=.33]{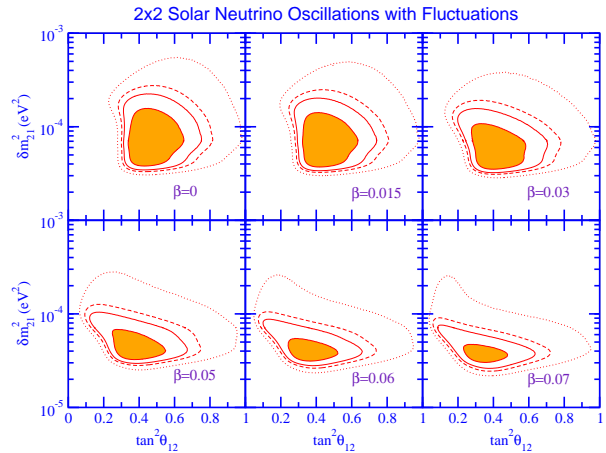}
\vspace*{+0cm} \caption{ \label{fig:4}  Allowed regions of the
  neutrino parameter space with solar-density fluctuations when only
  solar neutrino experiments (chlorine, all three gallium, SNO and SK
  experiments) are included in the analysis. The SSM density profile
  of Ref. \cite{Bahcall:2000nu} and the correlation length of 10 km
  are used. The case with no fluctuations ($\beta=0$) are compared
  with results obtained with the indicated fractional fluctuation.
  The shaded area is the 70 \% confidence level region. 90 \% (solid
  line), 95 \% (dashed line), and 99 \% (dotted line) confidence
  levels are also shown.}
\end{figure}

\Blue
\section{Results and Discussion}
\Black

In our analysis we used a covariance approach the details of which are
described in Ref. \cite{Balantekin:2003dc}. We use 93 data points in
our analysis; the  total rate of the chlorine experiment (Homestake
\cite{Cleveland:nv}), the average rate of the gallium experiments
(SAGE \cite{Abdurashitov:2002nt}, GALLEX \cite{Hampel:1998xg}, GNO
\cite{Altmann:2000ft}),  44 data points from the SK
zenith-angle-spectrum \cite{Fukuda:2001nj}, 34 data points from the
SNO day-night-spectrum \cite{Ahmad:2001an} and 13 data points from the
KamLAND spectrum \cite{Eguchi:2002dm}.  We took into account the
distribution of the neutrino sources in the Sun.  Specifically we
divided the Sun into several shells, calculated the survival
probability numerically for neutrino paths and averaged the survival
probabilities over the initial source distributions.  Similarly we
considered the effects of the matter density of the Earth (the
day-night effect) by solving neutrino evolution equations
numerically.  We illustrate the effects of source averaging (with and
without fluctuations) in Figures \ref{fig:2} and \ref{fig:3}. In
Figure \ref{fig:2} we show the source-averaged survival probability
separately for the pp and $^8$B neutrinos. In this figure the solid
line represents the survival probability of the neutrinos coming from
a single point (the center of the Sun) contrasted to the
source-averaged cases. For $\beta= 0 $, for a given $\delta m^2$,
neutrino energy, and mixing angle neutrino evolution is adiabatic for
the region of interest shown in the graph. In contrast, for $\beta
\neq 0$, one has a non-zero, but small hopping probability.  When the
neutrinos are created over a finite-size region (instead of a single
point) source-averaged survival probabilities do not coincide at large
$\delta m^2 / E$ values with the point-source survival probability
unlike the $\beta = 0$ case. Fluctuations also reduce the effect of
the averaging.  In Figure \ref{fig:3} we present the source-averaged
survival probabilities detected at the location of SNO for $^8$B
neutrinos with and without solar-density fluctuations. To calculate
this figure we used  the day and night live-time information from
Ref. \cite{HOWTOSNO}. One again observes that fluctuations smoothen
the survival probabilities.

\begin{figure}[t]
\vspace*{+.3cm}
\includegraphics[scale=.33]{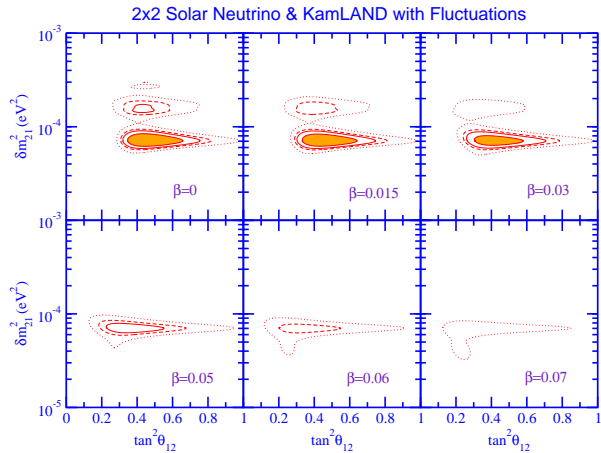}
\vspace*{+0cm} \caption{ \label{fig:5}  Allowed regions of the
  neutrino parameter space with solar-density fluctuations when the
  data from the   solar neutrino experiments (chlorine, all three
  gallium, SNO and SK experiments) and the KamLAND data are included
  in the analysis. The SSM density profile of
  Ref. \cite{Bahcall:2000nu} and the correlation length of 10 km are
  used. The case with no fluctuations ($\beta=0$) are compared with
  results obtained with the indicated fractional fluctuation.  The
  shaded area is the 70 \% confidence level region. 90 \% (solid
  line), 95 \% (dashed line), and 99 \% (dotted line) confidence
  levels are also shown.}
\end{figure}

\begin{figure}[t]
\vspace*{+.3cm}
\includegraphics[scale=.33]{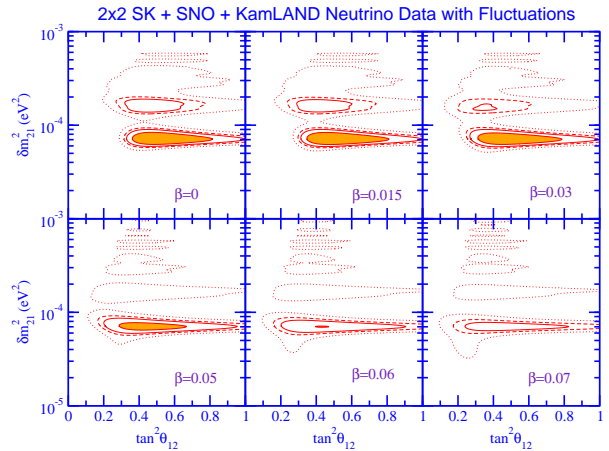}
\vspace*{+0cm} \caption{ \label{fig:6}  Allowed regions of the
  neutrino parameter space with solar-density fluctuations when the
  data from the solar neutrino experiments detecting only the high
  energy $^8$B neutrinos (SNO and SK experiments) and the KamLAND data
  are included in the analysis. The SSM density profile of
  Ref. \cite{Bahcall:2000nu} and the correlation length of 10 km are
  used. The case with no fluctuations ($\beta=0$) are compared with
  results obtained with the indicated fractional fluctuation.  The
  shaded area is the 70 \% confidence level region. 90 \% (solid
  line), 95 \% (dashed line), and 99 \% (dotted line) confidence
  levels are also shown.}
\end{figure}

We next turn our attention to parameter-space searches.  In
Ref. \cite{Balantekin:2003dc} from a global analysis of solar neutrino
and KamLAND data we found for electron neutrino oscillations into
another active flavor, the best fit values of  $\tan^2 \theta_{12}
\sim 0.46$ for the mixing angle between first and second generations,
$\tan^2 \theta_{13} \sim 0$ for the mixing between first and third
generations, and $\delta m_{21}^2 \sim 7.1 \times 10^{-5}$
eV$^2$. Other groups doing similar analyses found very similar best
fit values  \cite{allothers}. In Ref. \cite{Balantekin:2003dc} $\beta$
was taken to be zero.  We find that non-zero values of $\beta$ reduces
the size of the allowed region as shown in Figure \ref{fig:4}.  In
this figure allowed regions of the neutrino parameter space are shown
for different values of $\beta$ when all the solar neutrino
experiments (chlorine, all three gallium, SNO and SK experiments) are
included in the analysis. Although the minimum value of $\chi^2$ is
achieved when $\beta = 0$, one observes that for values of $\beta$ as
large as 0.07 allowed regions remain even at the 70 \% confidence
level.  Thus we conclude that solar neutrino data alone does not
significantly constrain the fluctuation parameter. Note that for
larger values of $\beta$ the region with larger values of $\delta m^2$
are no longer allowed. Since KamLAND data favor larger values of
$\delta m^2$,   incorporating KamLAND results dramatically reduces the
allowed region in parameter space as shown in  Figure
\ref{fig:5}. Again the  he minimum value of $\chi^2$ is achieved when
$\beta = 0$. However, in contrast to the calculation presented in
Figure \ref{fig:4}, one can put stringent limits on the amount of
fluctuation.  We find that $\beta < 0.05$ at the 70 \% confidence
level,  $\beta < 0.06$ at  the  at the 90 \% confidence level, and
$\beta < 0.07$ at  the  at the 95 \% confidence level.

Let us recall that both calculations are done for a value of $\tau =
10$ km for the  correlation length. Note that only the combination
$\beta^2 \tau$ enters into the calculation
(cf. Eq. (\ref{kformula}). Hence the $\beta$ values can be scaled by
adjusting the value of $\tau$ and for larger correlation lengths the
limits quoted above get tighter. However one cannot consider
arbitrarily large values of the correlation length. Our formulation of
the problem (the delta-function correlation approximation) becomes
unrealistic for larger values of $\tau$ as we illustrated in
Eq. (\ref{tau1}).  In we insert the best fit (minimum $\chi^2$) values
of  $\delta m^2 \sim 7.1 \times 10^{-5}$ eV$^2$ and $\tan^2 \theta
\sim 0.46$ into Eq. (\ref{tau1}) we find
\begin{equation}
\tau ({\rm km}) <  6 E ({\rm MeV}). 
\label{tau2}
\end{equation}
Hence for lower-energy (pp) neutrinos the reliable correlation lengths
are smaller then 10 km. However for higher energy ($^8$B) neutrinos
one can safely consider longer correlation lengths. For both SK and
SNO the energy threshold is $\sim 5$ MeV for which Eq. (\ref{tau2})
yields $ \tau < 30$ km. To explore this feature we repeat our analysis
considering only SK and SNO data together with the KamLAND data. We
show the allowed regions of the parameter space in Figure \ref{fig:6}.
In this figure for better comparison to  Figure \ref{fig:5}  we took
$\tau$ to be 10 km, however even $\tau=20$ km would be reasonable.
The best fit is still with $\beta=0$.   We find that $\beta < 0.07$ at
the 70 \% confidence level with $\tau = 10$ km. This limit would scale
down to $\beta < 0.07 / \sqrt{2}  \sim 0.05$ at the 70 \% confidence
level with $\tau = 20$ km. 

\begin{figure}[t]
\vspace*{+.3cm}
\includegraphics[scale=.5]{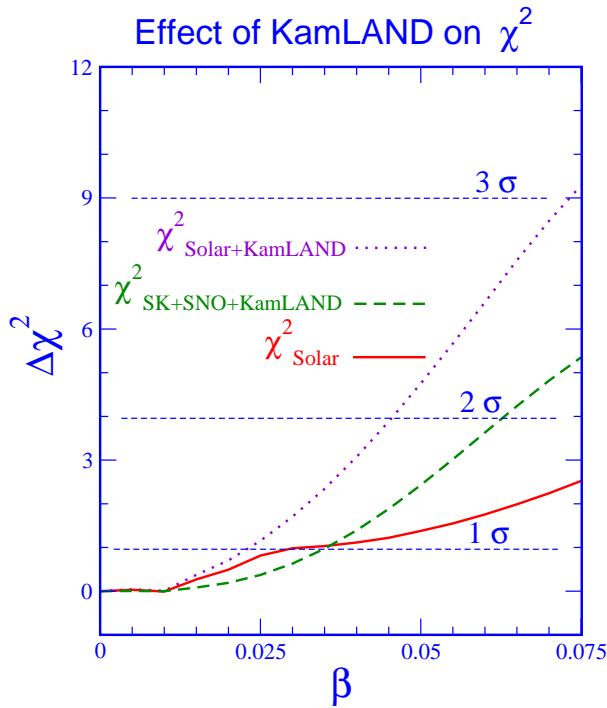}
\vspace*{+0cm} \caption{ \label{fig:7} Projection of the global
$\Delta \chi^2$ function on the fractional fluctuation parameter
$\beta$. The solid line represents the calculation using only the
solar neutrino data.  The dotted line is the calculation when the
KamLAND data are included along with the solar neutrino data. For
comparison we also show the calculation with solar-neutrino data
sensitive only to the $^8$B neutrinos (SK and SNO) combined with 
the KamLAND data.}
\end{figure}

In Figure \ref{fig:7} we present $\Delta \chi^2 = \chi^2 - \chi^2_{\rm
min}$  calculated as a function of $\beta$ when other parameters
$\delta m^2$ and  $\tan^2 \theta$ are unconstrained. In this figure
$\Delta \chi^2$ is projected only on one parameter ($\beta$) so that
$n-\sigma$ bounds on it are given by  $\Delta \chi^2 = n^2$. Clearly
KamLAND data plays a crucial role to constrain $\beta$. As the KamLAND
statistics improve in the near future we expect to improve our limits
on the fractional fluctuation.

In this paper we ignored fluctuations of another kind, namely magnetic
field fluctuations impacting neutrino evolution. It has been shown
that if neutrinos have sizable magnetic moments they can interact with
the transverse magnetic fields \cite{spinflavor1}  undergoing a
spin-flavor precession \cite{spinflavor2}. If the magnetic field is
noisy the spin-flavor precession will be impacted
\cite{Loreti:1994ry,Nicolaidis:da,Pastor:1995vn,Nunokawa:1997dp}.
However a relatively conservative assumption about the maximal size of
the solar magnetic field places the spin-flavor resonance at values of
$\delta m_{12}^2$ \cite{Balantekin:dv}
which is at the region of the neutrino parameter
space  ruled out by the KamLAND experiment. (For
an alternative approach see Ref.  \cite{Friedland:2002pg}). Hence one
can conclude that solar magnetic field fluctuations do not play a
role in the solar neutrino physics through the spin-flavor precession
mechanism.

We would like to point out that besides in the Sun (and other stars)
neutrinos interact with dense matter in several other sites such as
the early universe, supernovae, and newly-born neutron stars and
neutrino interactions with a stochastic background may play an even
more interesting role in those sites \cite{Prakash:2001rx}. Along
those lines a preliminary analysis of the effects of random density
fluctuations on matter enhanced neutrino flavor transitions in
core-collapse supernovae and implications of such fluctuations for
supernova dynamics and nucleosynthesis was given in
Ref. \cite{Loreti:1995ae}.

\Blue
\section*{ACKNOWLEDGMENTS}
\Black
This work was supported in part by the U.S. National Science
Foundation Grants No.\ PHY-0070161 and in part by
the University of Wisconsin Research Committee with funds granted by
the Wisconsin Alumni Research Foundation.

\Blue


\end{document}